\documentclass[prd,twocolumn,floatfix,amsmath,amssymb,floatfix]{revtex4}
\usepackage{graphicx,color,dcolumn,booktabs,bm}
\usepackage{longtable,lscape}
\usepackage{txfonts}
\usepackage{overpic}
\usepackage{amssymb}
\usepackage{indentfirst}
\usepackage{feynmf}   %{feynmp}
\usepackage{slashed}  %for Feynman symbols
\usepackage{cases}
\usepackage{color}
\usepackage{multirow}
\usepackage{epstopdf}
\usepackage{graphicx,color,dcolumn,booktabs,bm}

\def\Dsa{D_{s0}^{\ast+} (2317)}
\def\Dsb{D_{s1}^{+}(2460)}
\graphicspath{{Figures/}} %
\usepackage[colorlinks,
                      citecolor=blue,
                      anchorcolor=red,
                      menucolor=red,
                      linkcolor=red,
                      filecolor=red,
                      runcolor=red,
                      urlcolor=blue,
                      frenchlinks=red]{hyperref}
\begin{document}

\title{Pionic and radiative transitions from $T_{c\bar{s}0}^+(2900)$ to $D_{s1}^+(2460)$ as a probe of the structure of $D_{s1}^+(2460)$}
\author{Zi-Li Yue$^{1}$}
\author{Cheng-Jian Xiao$^2$}
\author{Dian-Yong Chen$^{1,3}$\footnote{Corresponding author}} \email{chendy@seu.edu.cn}
\affiliation{
 $^{1}$ School of Physics, Southeast University,  Nanjing 210094, China}
\affiliation{
$^{2}$ Institute of Applied Physics and Computational Mathematics, Beijing 100088, China} 
\affiliation{$^3$Lanzhou Center for Theoretical Physics, Lanzhou University, Lanzhou 730000, China}

\date{\today}

\begin{abstract}
In this work, we evaluated the widths of the pionic and radiative transitions from the $T_{c\bar{s}0}^{+}(2900)$ to the $D_{s1}^{+}(2460)$ in the $D_{s1}^{+}(2460)$ molecular frame and the $D_{s1}^{+}(2460)$ charmed-strange meson frame. Our estimations demonstrate that the transition widths in the $D_{s1}^{+}(2460)$ molecular frame are much larger than those in the the $D_{s1}^{+}(2460)$ charmed-strange meson frame. Specifically, the ratio of the widths of $\Gamma(T_{c\bar{s}0}^{+}(2900)\to D_{s1}^{+} \pi^{0})$ and $\Gamma(T_{c\bar{s}0}^{+}(2900)\to D^{+(0)}K^{0(+)})$ is estimated to be around 0.1 in the $D_{s1}^{+}(2460)$ charmed-strange meson frame, whereas the lower limit of this ratio is 0.67 in the $D_{s1}^{+}(2460)$ molecular frame. Thus, the aforementioned ratio could be employed as a tool for testing the nature of the $D_{s1}^{+}(2460)$.
\end{abstract}

%\pacs{}

\maketitle

%%%%%%%%%%%%%%%%%%%%%%%%%%%%%%%%%%
\section{Introduction}
\label{sec:introduction}

As one of typical new hadron states with one heavy quark, $D_{s0}^{\ast+}(2317)$ was first observed by BABAR Collaboration in the inclusive $D_{s}^{+}\pi^{0}$ invariant mass distribution from the electron-positron annihilation data at energies near 10.6 GeV~\cite{BaBar:2003oey}, and then confirmed by the CLEO Collaboration~\cite{CLEO:2003ggt}. As indicated in Ref.~\cite{BaBar:2003oey}, the most possible $J^P$ quantum numbers of $D_{s0}^{\ast+}(2317)$ were $0^+$, which could be a good candidate of $P$-wave charmed strange meson~\cite{Godfrey:2003kg,Bardeen:2003kt,Colangelo:2003vg,Fayyazuddin:2003aa,Colangelo:2005hv,Lu:2006ry,Song:2015nia}. However, the observed mass of $D_{s0}^{\ast+}(2317)$ is about 160 MeV below the corresponding predicted mass of the $P$-wave charmed strange meson, which is 2.48 MeV~\cite{Godfrey:1985xj,Godfrey:2003kg}. Moreover, the observed mass of $\Dsa$ is about 40 MeV below the threshold of $DK$, which indicates $\Dsa$ could be a good candidate of $DK$ molecular state~\cite{Faessler:2007gv,Xiao:2016hoa,Hwang:2004cd,Xie:2010zza,Zhang:2006ix,Chen:2004dy,Dong:2017gaw,Wei:2005ag}. 

Besides confirming the existence of $\Dsa$~\cite{CLEO:2003ggt}, the CLEO Collaboration reported another new narrow resonance $\Dsb$ in the invariant mass distribution of $D_s^{\ast+} \pi^0$ with the mass around 2.46 GeV, and the $J^{P}$ quantum numbers were determined to be $1^+$~\cite{CLEO:2003ggt}. Similar to the case of $\Dsa$, $\Dsb$ could be a candidate of $P$ wave charmed-strange mesons\cite{Godfrey:2003kg,Bardeen:2003kt,Colangelo:2003vg,Fayyazuddin:2003aa,Colangelo:2005hv,Lu:2006ry,Song:2015nia}. However, the mass of $\Dsb$ is also about 100 MeV below the predicted mass of the corresponding $P$-wave charmed strange meson~\cite{Godfrey:1985xj}, which makes the interpretation of $\Dsb$ in the conventional charmed strange frame questionable~\cite{Godfrey:1985xj,Rosner:2006vc}. It is more interesting to notice that the mass of $\Dsb$ is also about 40 MeV below the threshold of $D^\ast K$, which lead to the prosperity of $D^\ast K$ molecular interpretation~\cite{Xiao:2016hoa,Faessler:2007us,Tang:2023yls,Dong:2017gaw,Kolomeitsev:2003ac,Guo:2006fu,Close:2005se,Wei:2005ag}.
    
Recently, the LHCb Collaboration reported two new structures $T^{0/++}_{c\bar{s}0}(2900)$ in $D_{s}^+\pi^{+}/D_{s}^+\pi^{-}$ invariant mass spectrum of $B^+\to D^- D_s^+ \pi^+/B^0\to \bar{D}^0 D_s^+ \pi^-$ decays with a significance to be $9\sigma$. The masses and widths of the $T_{c\bar{s}0}^{0/++}(2900)$ are measured to be~\cite{LHCb:2022xob, LHCb:2022bkt},
\begin{eqnarray}
m_{T_{c\bar{s}0}^{0}}&=&2892\pm14\pm15\mathrm{MeV},\nonumber\\
\Gamma_{T_{c\bar{s}0}^{0}}&=&119\pm26\pm12\mathrm{MeV},
\end{eqnarray}
and
\begin{eqnarray}
m_{T_{c\bar{s}0}^{++}}&=&2921\pm17\pm19\mathrm{MeV},\nonumber\\
\Gamma_{T_{c\bar{s}0}^{++}}&=&137\pm32\pm14\mathrm{MeV},
\end{eqnarray}
respectively.

%\iffalse
\begin{figure}[t]
 \centering
  \includegraphics[width=8.5cm]{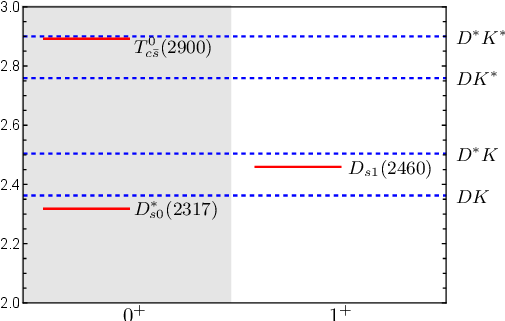}
  \caption{The masses of $D_{s0}^{*}(2317)$, $D_{s1}(2460)$, and~$T_{c\bar{s}0}^{0}$. The thresholds of $D^{(\ast)}K^{(\ast)}$ are also presented for comparison.}\label{Fig:Trim}
\end{figure}
%\fi

From the above parameters, one can conclude that $T_{c\bar{s}0}^0(2900)$ and $T^{++}_{c\bar{s}0}(2900)$ should be two of the isospin triplets. In addition, the experimental measurement indicates that the masses of $T_{c\bar{s}0}^{0/++}(2900)$ is near the threshold of $D^{*}K^{*}$, especially the neutral one. As shown in Fig.~\ref{Fig:Trim}, the newly observed $T_{c\bar{s}0}(2900)$, along with $D_{s0}^{*+}(2317)$ and $D^+_{s1}(2460)$,  make the states near the $D^{(*)}K^{(*)}$ thresholds abundant\cite{Xiao:2020ltm,Yue:2022gym,Belle:2021nuv,Duan:2021pll,BESIII:2013mhi,BaBar:2003oey,CLEO:2003ggt,Kim:2005gt,Nielsen:2005ia,Rosner:2006vc,Zhang:2006ix,Wu:2021udi,Wu:2019vbk,Ji:2022uie,Fu:2021wde,Ortega:2023azl}. However, unlike the $D_{s0}^{*+}(2317)$ and $D_{s1}^{+}(2460)$, the isospin of $T_{c\bar{s}0}(2900)$ is $1$, which suggests that it could only be an exotic candidate rather than conventional charmed strange meson. Consequently, certain exotic interpretations, particularly the $D^\ast K^\ast$ molecular interpretation, have been proposed\cite{Chen:2017rhl,Liu:2022hbk,Wei:2022wtr,Molina:2022jcd,Ge:2022dsp,Chen:2022svh,Agaev:2022duz,Duan:2023qsg}. For instance, the authors in Ref.~\cite{Chen:2022svh} suggested that the $T_{c\bar{s}0}^{0/++}(2900)$ could be a $D^{*}K^{*}$ molecular state with $I(J^{P})=1(0^{+})$, employing the one-boson exchange model. In our previous work~\cite{Yue:2022mnf}, we examined the strong decay behavior of $T_{c\bar{s}0}^{0}(2900)$ in $D^{*}K^{*}$ molecular scenario using an effective Lagrangian approach. Specifically, we investigated the decay $T_{c\bar{s}0}^{0}(2900) \to D_{s1}^{+}(2460) \pi^{-} $ where $D_{s1}^{+}(2460)$ is considered to be a $P$ wave charmed-strange meson. In the present work, we investigate the pionic and radiative transitions from $T_{c\bar{s}0}^{+}(2900)$ to $D_{s1}^{+}(2460)$ within a molecular framework, where $T_{c\bar{s}0}^{+}(2900)$ and $D_{s1}^{+}(2460)$ are assigned to be the $D^{*}K^{*}$ and $D^{\ast} K$ molecules, respectively. By comparing the results in the molecular scenario with those in the $P$ wave charmed strange meson scheme, we demonstrate that the pionic and radiative transition process explored in the present work may be utilized to probe the nature of $D_{s1}^{+}(2460)$.

This work is organized as follows. After introduction, the hadronic molecular structure of $T_{c\bar{s}0}^{+}(2900)$ and $D_{s1}^{+}(2460)$ are discussed in Section \ref{Sec:Structure}, The pionic and radiative transitions between $T_{c\bar{s}0}^{+}(2900)$ and $D_{s1}^{+}(2460)$ are presented in Section \ref{Sec:Transition}. The numerical results and related discussions are presented in Section \ref{Sec:Results} and the last section is devoted to a short summary.

\section{Hadronic molecular structure of $T_{c\bar{s}0}^{+}(2900)$ and $D_{s1}^{+}(2460)$}
\label{Sec:Structure}

In the molecular scheme, the $T_{c\bar{s}0}^{+}(2900)$ and $D_{s1}^{+}(2460)$ could be considered as $S$-wave molecular states composed of $D^{*}K^{*}$ and $D^{*}K$, respectively. Here, we employ the effective Lagrangian approach to describe the coupling of the molecular states with their components, and the effective Lagrangians related to $T_{c\bar{s}0}^{+}(2900)$ and $D_{s1}^{+}(2460)$ are,

\begin{widetext}
	\begin{eqnarray}
\mathcal{L}_{D_{s1}}&=&g_{D_{s1}}D_{s1}^{\mu+}(x)\int dy\Phi_{D_{s1}}(y^{2})\Big[D^{*+}_{\mu}(x+\omega_{KD^{*}}y)K^0(x-\omega_{D^{*}K}y)+D^{*0}_{\mu}(x+\omega_{KD^{*}}y)K^+(x-\omega_{D^{*}K}y)\Big],\nonumber\\
\mathcal{L}_{T_{c\bar{s}0}}&=&g_{T_{c\bar{s}0}}T_{c\bar{s}0}^{+}(x)\int dy\Phi_{T_{c\bar{s}0}}(y^{2})\Big[D^{*+}_{\mu}(x+\omega_{K^{*}D^{*}}y)K^{*0\mu}(x-\omega_{D^{*}K^{*}}y)-D^{*0}_{\mu}(x+\omega_{K^{*}D^{*}}y)K^{*+\mu}(x-\omega_{D^{*}K^{*}}y)\Big]\label{Eq:1},
\end{eqnarray}
\end{widetext}
respectively, where $\omega_{ij}=m_{i}/(m_{i}+m_{j})$ is the kinematical parameter. The $\Phi_{T_{c\bar{s}0}}(y^{2})$ and $\Phi_{D_{s1}}(y^{2})$ are the correlation function for $T_{c\bar{s}0}^{+}(2900)$ and $\Dsb$, respectively, which are introduced to describe the molecular inner structure. The Fourier transformation of the correlation function is,
\begin{eqnarray}
\Phi_{M}(y^{2})=\int \frac{d^{4}p}{(2\pi)^{4}}e^{-ipy}\tilde{\Phi}_{X}(-p^{2},\Lambda^{2}_{M}),\quad M=\Big(T_{c\bar{s}0},&D_{s1} \Big).\quad
\end{eqnarray}

In principle, the correlation function in momentum space should decrease sharply enough to avoid the divergence in the ultraviolet region. Here, we employ the correlation function in the Gaussian form~\cite{Chen:2015igx,Chen:2013cpa,Dong:2013kta,Xiao:2020alj}, which is,
\begin{eqnarray}
\tilde{\Phi}_{M}(p_{E}^{2},\Lambda_{M}^{2})=\mathrm{exp}(-p_{E}^{2}/\Lambda_{M}^{2})
\end{eqnarray}
where $P_{E}$ is the Jacobi momentum in the Euclidean space, and $\Lambda_{M}$ is a model parameter to depict the distribution of components in the molecule.

For the coupling constants $g_{T_{c\bar{s}0}}$ and $g_{D_{s1}}$ in Eq.~(\ref{Eq:1}), they could be determined by the Weinberg's compositeness condition, which means that the possibility of finding the molecular in a bare elementary state is set equal to zero~\cite{vanKolck:2022lqz,Weinberg:1962hj,Salam:1962ap,Ivanov:1996fj,Ivanov:1999bk}, i.e.,
\begin{eqnarray}
Z_{T_{c\bar{s}0}}&=&1-\Pi^{\prime}_{T_{c\bar{s}0}}=0\nonumber\\
Z_{D_{s1}}&=&1-\Pi^{\prime}_{D_{s1}}=0\label{Eq:2}
\end{eqnarray}
with $\Pi^{\prime}_{T_{c\bar{s}0}}$ to be the derivative of mass operator of the $T_{c\bar{s}0}$. While for $\Dsb$, the mass operator $\Pi_{D_{s1}}^{\mu\nu}$ could be divided into the transverse part $\Pi_{D_{s1}}$ and the longitudinal part $\Pi_{D_{s1}}^{L}$, which is,
\begin{eqnarray}
\Pi^{\mu\nu}_{D_{s1}}(p)=g^{\mu\nu}_{\bot}\Pi_{D_{s1}}(p^{2})+\frac{p^{\mu}p^{\nu}}{p^{2}}\Pi^{L}_{D_{s1}}(p^{2})
\end{eqnarray}
where $g^{\mu\nu}_{\bot}=g^{\mu\nu}-p^{\mu}p^{\nu}/p^{2}$, $g^{\mu\nu}_{\bot}p_{\mu}=0$. 

\begin{figure}[t]
\begin{tabular}{cc}
  \centering
  \includegraphics[width=4.2cm]{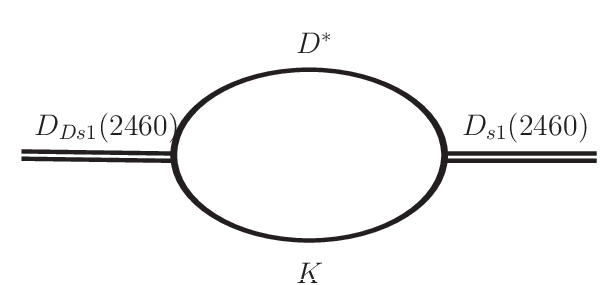}&
  \includegraphics[width=4.2cm]{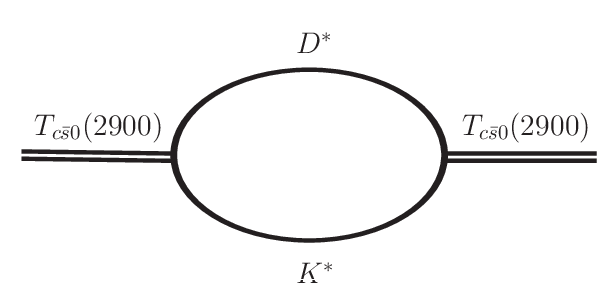}\\
\\
 $(a)$ & $(b)$ \\
 \end{tabular}
\caption{Mass operators of the $D_{s1}^+(2460)$ (diagram (a))~and~$T_{c\bar{s}0}^+(2900)$ (diagram (b)).}\label{Fig:Tri1}
\end{figure}

According to the effective Lagrangian shown in Eq.~(\ref{Eq:1}), the concrete forms of mass operator of the $T_{c\bar{s}0}$ and $D_{s1}$ corresponding to Fig.~\ref{Fig:Tri1}-(a) and~\ref{Fig:Tri1}-(b) are,
\begin{eqnarray}
\Pi^{\mu\nu}(m_{D_{s1}}^{2})&=& g^2_{D_{s1}}\int \frac{d^{4}q}{(2\pi)^{4}}\tilde{\Phi}_{D_{s1}}^{2}[-(q-\omega_{D^{*}K}p)^{2},\Lambda_{D_{s1}}^{2}]\nonumber\\&\times&\frac{1}{(p-q)^{2}-m_{K^{*}}^{2}}\frac{-g^{\mu\nu}+q^{\mu}q^{\nu}/m_{D^{*}}^{2}}{q^{2}-m_{D^{*}}^{2}}\nonumber\\
\Pi ({m_{T_{c\bar{s}0}}^{2}})&=& g^2_{T_{c\bar{s}0}}\int \frac{d^{4}q}{(2\pi)^{4}}\tilde{\Phi}_{T_{c\bar{s}0}}^{2}[-(q-\omega_{D^{*}K^{*}}p)^{2},\Lambda_{T_{c\bar{s}0}}^{2}]\nonumber\\&\times&\frac{-g^{\mu\nu}+q^{\mu}q^{\nu}/m_{D^{*}}^{2}}{q^{2}-m_{D^{*}}^{2}}\frac{-g^{\mu\nu}+(p-q)^{\mu}(p-q)^{\nu}/m_{{K}^{*}}^{2}}{(p-q)^{2}-m_{K^{*}}^{2}},\nonumber\\ 
\label{Eq:MO}
\end{eqnarray}
respectively.

\section{Pionic and radiative transitions from $T_{c\bar{s}0}^{+}(2900)$ to $D_{s1}^{+}(2460)$}
\label{Sec:Transition}
\begin{figure}[t]
\begin{tabular}{cc}
  \centering
  \includegraphics[width=4.2cm]{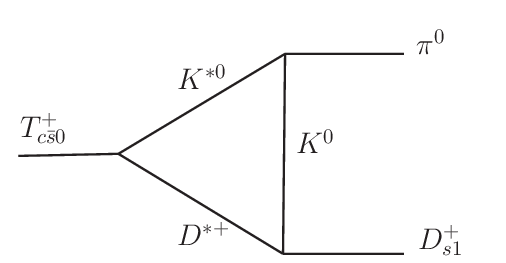}&
 \includegraphics[width=4.2cm]{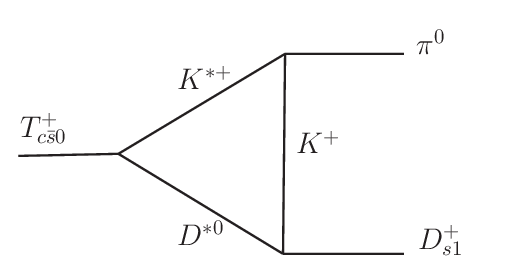}\\
\\
 $(a)$ & $(b)$ \\
 \end{tabular}
\caption{Diagrams contributing to the pionic transition from the $T_{c\bar{s}0}^{+}(2900)$ to $D_{s1}^{+}(2460)$.}\label{Fig:Tri2}
\end{figure}

\begin{figure}[t]
\begin{tabular}{cc}
  \centering
  \includegraphics[width=4.2cm]{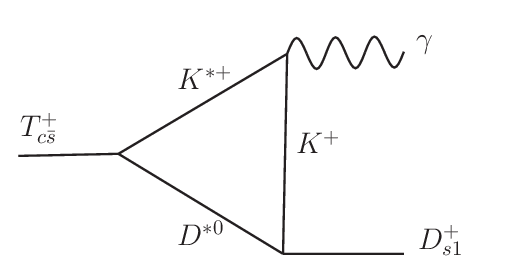}&
 \includegraphics[width=4.2cm]{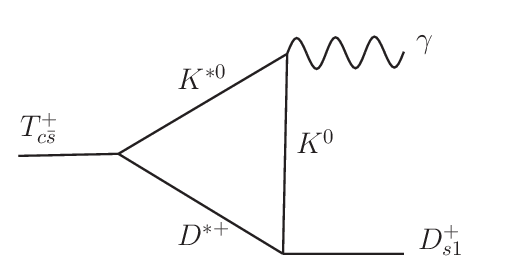}\\
\\
 $(c)$ & $(d)$ \\
 \end{tabular}
\caption{Diagrams contributing to the radiative transition from the $T_{c\bar{s}0}^{+}(2900)$ to $D_{s1}^{+}(2460)$.}\label{Fig:Tri3}
\end{figure}

In the present work, the initial state $T_{c\bar{s}0}^{+}(2900)$ is considered as a $D^\ast K^\ast$ molecule. Subsequently, the pionic and radiative transitions from $T_{c\bar{s}0}^{+}(2900)$ to $D_{s1}^{+}(2460)$ could occur through two possible subprocesses. The first one is via the subprocess $K^\ast \to K \pi/\gamma$ and the $K$ and $D^\ast$ couple to the $\Dsb$, which is shown in Figs.~\ref{Fig:Tri2}-\ref{Fig:Tri3}. The second one is through the subprocess $D^\ast \to D \pi /\gamma$ and the $D$ and $K^\ast$ couple to the $\Dsb$, where the exchanged meson is $D$ meson. The mass of $D$ meson is much greater than the one of $K$ meson, thus, the contributions from the second subprocess should be suppressed. In addition, in the kaon exchange  diagram, the final $D_{s1}(2460)$ couples to $DK$, and  the threshold of $DK$ is close to the mass of the $D_{s1}(2460)$, thus in the triangle diagram, all the involved internal particles are almost on-shell, which will enhance the loop integral. On the contrary, in the $D$ meson exchange diagram, the final $D_{s1}(2460)$ couples to $D K^\ast$, while the threshold of $D K^\ast$ is far above the mass of $D_{s1}(2460)$, thus, the involve internal particles are off-shell, which further suppress the contributions from the $D$ meson exchange diagrams. Thus in the present estimation, we only consider the diagrams in Figs.~\ref{Fig:Tri2}-\ref{Fig:Tri3}.

In the present calculations, the diagrams in Figs.~\ref{Fig:Tri2}-\ref{Fig:Tri3} are evaluated in the hadronic level. The interactions of the involved particles are depicted by effective Lagrangian. The effective Lagrangian depicts the subprocess $K^\ast \to K\pi$ could be constructed by SU$(3)$ symmetry interaction~\cite{Xiao:2016hoa,Yue:2022mnf,Xiao:2020ltm,Oh:2000qr,Haglin:2000ar,Chen:2010re,Dong:2008gb}, which is
\begin{eqnarray}
\mathcal{L}_{K^{*} K \pi }&=&-i g_{K^{*} K \pi}\left(\bar{K} \partial^{\mu} \pi-\partial^{\mu} \bar{K} \pi\right) K_{\mu}^{*}+\text { H.c. },
\end{eqnarray}
while the one for $K^\ast \to K\gamma$ is,
\begin{eqnarray}
\mathcal{L}_{K^{*} K \gamma}&=&\left( \frac{g_{K^{*+} K^{+} \gamma}}{4} e \epsilon^{\mu \nu \alpha \beta} F_{\mu \nu} K_{\alpha \beta}^{*+} K^{-}\right. \nonumber\\
&+&\left.\frac{g_{K^{* 0} K^{0} \gamma}}{4} e \epsilon^{\mu \nu \alpha \beta} F_{\mu \nu} K_{\alpha \beta}^{* 0} \bar{K}^{0}\right)+\text { H.c. },
\end{eqnarray}
where $F_{\mu\nu}=\partial_{\mu}A_{\nu}-\partial_{\nu}A_{\mu}$, $K_{\alpha\beta}^{*}=\partial_{\alpha}K_{\beta}^{*}-\partial_{\beta}K_{\alpha}^{*}$ are the field-strength tensors. According to the decay width of $K^\ast K \pi$~\cite{ParticleDataGroup:2022pth}, the  coupling constant $g_{K^{*}K\pi}$ is estimated to be 3.12. Additionally, we utilize the coupling constants $g_{K^{*0}K^{0}\gamma}=-1.27$ and $g_{K^{*+}K^{+}\gamma}=0.83$, which are  estimated from the corresponding partial width of $K^{*+/0}\to K^{+/0}\gamma$~\cite{ParticleDataGroup:2022pth,ParticleDataGroup:2008zun,Chen:2010re,Dong:2008gb}.


\subsection{In the $\Dsb$ molecular frame}
With the above preparation, we can estimate the pionic and radiative transitions from $T_{c\bar{s}0}^{+}(2900)$ to $\Dsb$ in a molecular frame, where both $T_{c\bar{s}0}^{+}(2900)$ and $\Dsb$ are considered as molecular states. The amplitude corresponding to Fig.~\ref{Fig:Tri2}-(a) is,
\begin{eqnarray}
i\mathcal{M}_{a}&=&i^{3}\int \frac{d^{4}q}{(2\pi)^{4}}\left[g_{T_{c\bar{s}0}}\tilde{\Phi}_{T_{c\bar{s}0}}(-p_{12}^{2},\Lambda^{2}_{T_{c\bar{s}0}})g_{\mu\alpha}\right]\nonumber\\&\times&\left[g_{D_{s1}}\tilde{\Phi}_{D_{s1}}(-p_{20}^{2},\Lambda^{2}_{D_{s1}})\epsilon_{\beta}(p_{4})\right]\nonumber\\&\times&\left[-ig_{K^{*}KP}i(p_{3\nu}-g_{\nu})\right]\frac{-g^{\mu\nu}+p_{1}^{\mu}p_{1}^{\nu}/m_{1}^{2}}{p_{1}^{2}-m_{1}^{2}}\nonumber\\&\times&\frac{-g^{\alpha\beta}+p_{2}^{\alpha}p_{2}^{\beta}/m_{2}^{2}}{p_{2}^{2}-m_{2}^{2}}\frac{1}{q^{2}-m_{q}^{2}}\mathcal{F}^{2}(m_{q},\Lambda) \label{Eq:Ma}
\end{eqnarray}
where~$p_{12}=p_{1}\omega_{D^{*}K^{*}}-p_{2}\omega_{K^{*}D^{*}}$~and~$p_{20}=p_{2}\omega_{KD^{*}}-q\omega_{D^{*}K}$. The amplitudes corresponding to Fig.~\ref{Fig:Tri2}-(b) can be obtained by $\mathcal{M}_{a}$ through replacing the masses of the involve meson and the relevant coupling constants with the corresponding values, which is,
\begin{eqnarray}
i\mathcal{M}_{b}=\left.i\mathcal{M}_{a}\right|
_{m_{K^{*0}}\to m_{K^{*+}}, m_{D^{*+}}\to m_{D^{*0}}, m_{K^{0}}\to m_{K^{+}}}^{g_{K^{*+}K^{+}\pi^{0}}\to g_{K^{*0}K^{0}\pi^{+}}}.
\end{eqnarray}
Then, the total amplitude of $T_{c\bar{s}0}^{+}(2900)\to \Dsb \pi^0$ is,
\begin{eqnarray}
i\mathcal{M}_{T_{c\bar{s}0}\to D_{s1}\pi}=i\mathcal{M}_{a}+i\mathcal{M}_{b}
\end{eqnarray}

In the similar way, one can obtain the amplitudes for $T_{c\bar{s}0}^{+}(2900)\to \Dsb \gamma$ corresponding to the diagrams in Fig.~\ref{Fig:Tri3}, which are,
\begin{eqnarray}
i\mathcal{M}_{c}&=&i^{3}\int \frac{d^{4}q}{(2\pi)^{4}}\left[g_{T_{c\bar{s}0}}\tilde{\Phi}_{T_{c\bar{s}0}}(-p_{12}^{2},\Lambda^{2}_{T_{c\bar{s}0}})g_{\phi\tau}\right]\nonumber\\&\times&\left[g_{D_{s1}}\tilde{\Phi}_{D_{s1}}(-p_{20}^{2},\Lambda^{2}_{D_{s1}})\epsilon_{o}(p_{4})\right]\left[-\frac{g_{K^{*}K\gamma}}{4}e\epsilon^{\mu\nu\alpha\beta}\right.\nonumber\\&\times&\left.(ip_{3\mu}g_{\nu\theta}-ip_{3\nu}g_{\mu\theta})(-ip_{1\alpha}g_{\beta\delta}+ip_{1\beta}g_{\alpha\delta})\right.\nonumber\\&\times&\left.\epsilon^{\theta}(p_{3})\right]\frac{-g^{\phi\delta}+p_{1}^{\phi}p_{1}^{\delta}/m_{1}^{2}}{p_{1}^{2}-m_{1}^{2}}\frac{-g^{\tau o}+p_{2}^{\tau}p_{2}^{o}/m_{2}^{2}}{p_{2}^{2}-m_{2}^{2}}\nonumber\\&\times&\frac{1}{q^{2}-m_{q}^{2}}\mathcal{F}^{2}(m_{q},\Lambda),\nonumber\\
i\mathcal{M}_{d}&=&\left.i\mathcal{M}_{c}\right|
_{m_{K^{*0}}\to m_{K^{*+}}, m_{D^{*+}}\to m_{D^{*0}}, m_{K^{0}}\to m_{K^{+}}}^{g_{K^{*+}K^{+}\gamma}\to g_{K^{*0}K^{0}\gamma}}, \label{Eq:Mc}
\end{eqnarray}
then, the total amplitude of $T_{c\bar{s}0}^{+}(2900)\to \Dsb \gamma$ is,
\begin{eqnarray}
i\mathcal{M}_{T_{c\bar{s}0}\to D_{s1}\gamma}=i\mathcal{M}_{c}+i\mathcal{M}_{d}.\label{Eq:RadTot1}
\end{eqnarray}
here, we introduce a monopole form factor to depict the inner configuration and off-shell effect of the exchanging mesons,
\begin{eqnarray}
\mathcal{F}(m_{q},\Lambda)=\frac{m_{q}^{2}-\Lambda^{2}}{q^{2}-\Lambda^{2}},
\end{eqnarray}

After performing the loop integral, we find the amplitude of $T_{c\bar{s}0}^{+}(2900)\to \Dsb \gamma$ can be simplified as,
\begin{eqnarray}
\mathcal{M}_{T_{c\bar{s}0}\to D_{s1}\gamma}=g_{T_{c\bar{s}0}\to D_{s1}\gamma}\epsilon_{\mu\nu\alpha\beta}p_{3}^{\mu}p_{4}^{\nu}\epsilon^{\alpha}(p_{3})\epsilon^{\beta}(p_{4}),
\end{eqnarray}
which satisfies the principle of gauge invariance for the photon field.

\subsection{In the $\Dsb$ charmed-strange meson frame}
In the $\Dsb$ charmed-strange meson frame, the $\Dsb$ is considered as a $P$-wave charmed-strange meson. As for the interaction related to $P$-wave and $S$-wave charmed-strange mesons, the effective Lagrangian could be constructed by using heavy quark limit and chiral symmetry~\cite{He:2019csk,Casalbuoni:1996pg,Wise:1992hn,Yan:1992gz,Cheng:1992xi,Ding:2008gr}, which is 
\begin{eqnarray}
\mathcal{L}_{\mathcal{D}_{1} \mathcal{D}^{*} \mathcal{P}}&=& ig_{\mathcal{D}_{1} \mathcal{D}^{*} \mathcal{P}}\left(\mathcal{D}_{1}^{\mu}\stackrel{\leftrightarrow}{\partial}_{\nu}D^{*\dagger}_{\mu}\right)\partial^{\nu}\mathcal{P}+\rm{H.c.}, \label{Eq:LagD1DStarP}
\end{eqnarray}
the relevant coupling constant $g_{\mathcal{D}_{1} \mathcal{D}^{*} \mathcal{P}}$ is,
\begin{eqnarray}
g_{\mathcal{D}_{1}\mathcal{D}^{*}P}&=&-\frac{h}{f_{\pi}},
\end{eqnarray}
with $f_{\pi}=132~\rm{MeV}$ to be the decay constant of $\pi$ meson, and $h=0.56\pm0.04$~\cite{He:2019csk,Falk:1992cx,Liu:2020ruo,Chen:2019asm,Colangelo:2012xi}.

In our previous work~\cite{Yue:2022mnf}, we have estimated the pionic transition from $T_{c\bar{s}0}^{+}(2900)$ to $\Dsb$ in the $\Dsb$ charmed-strange meson frame. Thus, in this subsection, we only present the amplitude for $T_{c\bar{s}0}^{+}(2900)\to \Dsb\gamma$. The amplitudes corresponding to the diagrams in Fig.~\ref{Fig:Tri3} are,
\begin{eqnarray}
i\mathcal{M}^{\prime}_{c}&=&i^{3}\int \frac{d^{4}q}{(2\pi)^{4}}\left[g_{T_{c\bar{s}0}}\tilde{\Phi}(-p_{12}^{2},\Lambda^{2}_{T_{c\bar{s}0}})g_{\phi\tau}\right]\nonumber\\&\times&\left[-ig_{D_{1}D^{*}P}(p_{2}^{\lambda}+p_{4}^{\lambda})g_{\lambda}g^{o\rho}\epsilon^{\rho}(p_{4})\right]\nonumber\\&\times&\left[-\frac{g_{K^{*}K\gamma}}{4}e\epsilon^{\mu\nu\alpha\beta}i(p_{3\mu}g_{\nu\theta}-p_{3\nu}g_{\mu\theta})(-i)\right.\nonumber\\&\times&\left.(p_{1\alpha}g_{\beta\delta}-p_{1\beta}g_{\alpha\delta})\epsilon^{\theta}(p_{3})\right]\frac{-g^{\phi\delta}+p_{1}^{\phi}p_{1}^{\delta}/m_{1}^{2}}{p_{1}^{2}-m_{1}^{2}}\nonumber\\&\times&\frac{-g^{\tau o}+p_{2}^{\tau}p_{2}^{o}/m_{2}^{2}}{p_{2}^{2}-m_{2}^{2}}\frac{1}{q^{2}-m_{q}^{2}}\mathcal{F}^{2}(m_{q},\Lambda)\nonumber\\
i\mathcal{M}^{\prime}_{d}&=&\left.i\mathcal{M^{\prime}}_{c}\right|
_{m_{K^{*0}}\to m_{K^{*+}}, m_{D^{*+}}\to m_{D^{*0}}, m_{K^{0}}\to m_{K^{+}}}^{g_{K^{*+}K^{+}\gamma}\to g_{K^{*0}K^{0}\gamma}}. \label{Eq:Mcp}
\end{eqnarray}
The total amplitude for $T_{c\bar{s}0}^{+}(2900)\to \Dsb\gamma$ is,
\begin{eqnarray}
i\mathcal{M}^{\prime}_{T_{c\bar{s}0}\to D_{s1}\gamma}=i\mathcal{M}^{\prime}_{c}+i\mathcal{M}^{\prime}_{d}.
\end{eqnarray}
Similar to Eq.~(\ref{Eq:RadTot1}), the above amplitude also satisfies the principle of gauge invariance for the photon field.

Additionally, in the above amplitudes, a phenomenological form factor in monopole form is introduced, which can describe the inner structure and off-shell effect of the exchanging mesons. The concrete form of the form factor is,
\begin{eqnarray}
\mathcal{F}(m_{q},\Lambda)=\frac{m_{q}^{2}-\Lambda^{2}}{q^2-\Lambda^{2}},
\end{eqnarray}
with $\Lambda$ to be a model parameter, which should be of the order of unity.

\section{Numerical Results and Discussions}
\label{Sec:Results}

In the present estimations, the loop integrals in the mass operators and the amplitudes can be estimated by Schwinger parameterizations ~\cite{Schwinger:1951nm}. This parameterization scheme is more convenient to handle the four-momentum integrals with the correlation functions in the Gaussian form. 

\subsection{Coupling constants}

\begin{figure}[t]
  \includegraphics[width=8.cm]{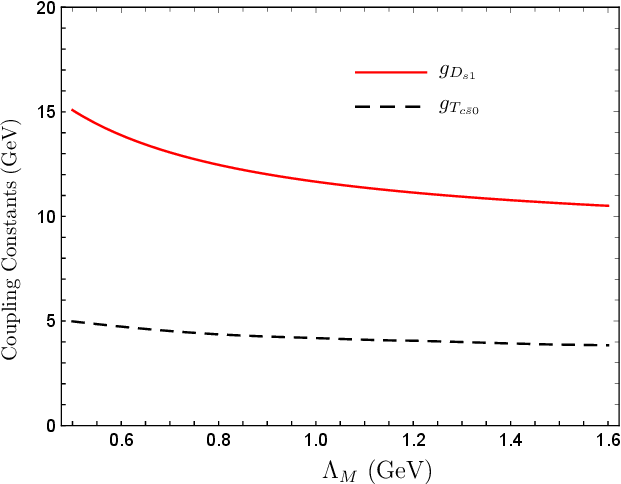}
\caption{The coupling constants $g_{T_{c\bar{s}0}}$ and $g_{D_{s1}}$ depending on model parameter $\Lambda_M$, where $\Lambda_{T_{c\bar{s}0}}=\Lambda_{D_{s1}}=\Lambda_M$.}\label{Fig:Tri4}
\end{figure}

Since the $T_{c\bar{s}0}^+(2900)$ has not been observed yet, in the present estimation, we take the same resonance parameters as those of $T_{c\bar{s}0}^0(2900)$ for easy comparison with our previous work in Ref.~\cite{Yue:2022gym}. Besides the coupling constants discussed in the above section, the coupling constants related to the molecular states are also needed to estimate the relevant decay widths. In the present work, we estimate the coupling constants $g_{D_{s1}}$ and $g_{T_{c\bar{s}0}}$ according to the compositeness conditions as shown in Eq.~(\ref{Eq:2}). Here, $\Lambda_{T_{c\bar{s}0}}$ and $\Lambda_{D_{s1}}$ are phenomenological model parameters, which should be of order $1~\rm{GeV}$. Considering both $K$ and $K^\ast$ are $S$-wave strange mesons and the similarity between $\Dsb$ and $T_{c\bar{s}0}^+(2900)$, we take $\Lambda_{T_{c\bar{s}0}}=\Lambda_{D_{s1}}=\Lambda_M$ for simplify. In Ref.~\cite{Yue:2022mnf}, our estimations indicate that the total width of $T_{c\bar{s}0}^{0}(2900)$ could be well reproduced with $\Lambda_M<1.6$, thus, in the present work, we vary the parameter $\Lambda_M$ from $0.5$ to $1.6~\rm{GeV}$, and the $\Lambda_M$ dependences of $g_{D_{s1}}$ and $g_{T_{c\bar{s}0}}$ are presented in Fig.~\ref{Fig:Tri4}. In the considered parameter range, one can find the coupling constants $g_{D_{s1}}$ and $g_{T_{c\bar{s}0}}$ are weakly dependent on the model parameter $\Lambda_M$. Particularly, with the parameter $\Lambda_M$ increase from $0.5$ to $1.2~\rm{GeV}$, the coupling constants $g_{T_{c\bar{s}0}}$ and $g_{D_{s1}}$ decrease from $4.98$ to $3.84~\rm{GeV}$ and from $15.08$ to $10.51~\rm{GeV}$, respectively.

\begin{figure}[t]
  \includegraphics[width=8.cm]{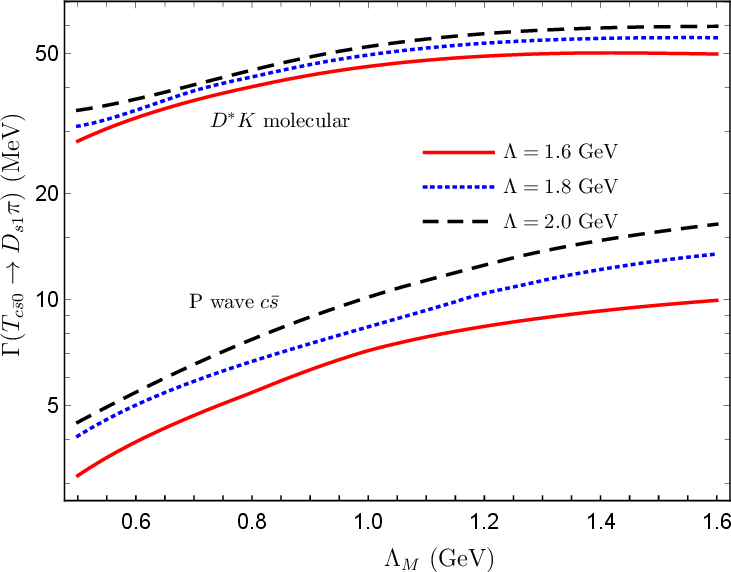}
\caption{The decay width of $T_{c\bar{s}0}^+ (2900)\to D_{s1}^+(2460)\pi^0$ in the $D_{s1}^+(2460)$ molecular scenario and in the $D_{s1}^+(2460)$ charmed strange meson frame, respectively.}\label{Fig:Pi}
\end{figure}

\begin{figure}[t]
  \includegraphics[width=8.cm]{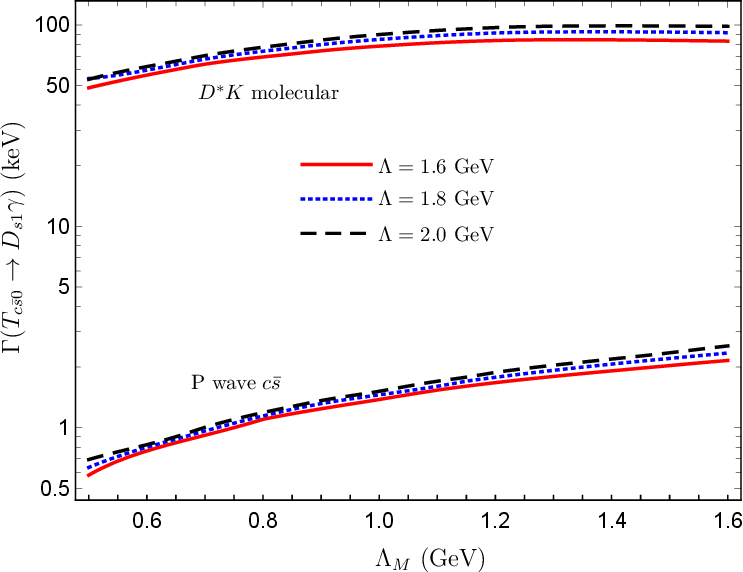}
\caption{The same as Fig.~\ref{Fig:Pi} but for $T_{c\bar{s}0}^+(2900)\to D_{s1}^+(2460)\gamma$. \label{Fig:Gamma}}
\end{figure}

\begin{table*}
\caption{The transition widths of $T_{c\bar{s}0}^+(2900) \to D_{s1}^+ (2460) \pi^0 $ in the $D_{s1}^+ (2460) $ charmed strange meson frame and the $D_{s1}^+ (2460) $  molecular state frame.
\label{Tab:Width}}
 \renewcommand\arraystretch{1.5}
\begin{tabular}{p{3.0cm}<\centering p{3cm}<\centering p{3.0cm}<\centering p{3.0cm}<\centering p{3.0cm}<\centering }
\toprule[1pt]
\multicolumn{5}{c}{Charmed-strange meson frame~\cite{Yue:2022mnf}}\\
%\midrule[1pt]
$\Lambda$& $\Lambda_M$ Range (GeV) & $\Gamma_{D_{s1}\pi}$ (MeV) & $\Gamma_{DK}$ (MeV) & $\Gamma_{D_{s1}\pi}/\Gamma_{DK}$ \\
\midrule[1pt]
1.6 & $1.03 \sim 1.56 $  &$7.36 \sim 9.81$&$53.05\sim101.06$&$0.10\sim0.14$\\ 
1.8 & $0.83 \sim 1.19 $  &$6.9\sim10.29$&$52.62\sim101.07$&$0.10\sim0.13$\\
2.0 & $0.71 \sim 1.01 $  &$6.63\sim10.26$&$53.37\sim100.28$&$0.10\sim0.12$\\
\midrule[1pt]
\multicolumn{5}{c}{Molecular state frame}\\
$\Lambda$& $\Lambda_M$ Range (GeV) & $\Gamma_{D_{s1}\pi}$ (MeV) & $\Gamma_{DK}$ (MeV) ~\cite{Yue:2022mnf}& $\Gamma_{D_{s1}\pi}/\Gamma_{DK}$ \\
\midrule[1pt]
1.6&$0.69\sim1.18$&$36.36\sim48.85$&$27.37\sim66.42$&$0.74\sim1.33$\\
1.8&$0.59\sim0.94$&$33.99\sim54.94$&$29.02\sim66.09$&$0.83\sim1.17$\\
2.0&$0.52\sim0.83$&$34.84\sim46.04$&$29.08\sim69.05$&$0.67\sim1.19$\\

\bottomrule[1pt]
\end{tabular}
 
\end{table*}

\subsection{Decay widths}

In addition to the parameter $\Lambda_M$ introduced by the correlation functions for $T_{c\bar{s}0}^{+}(2900)$ and $D_{s1}^{+}(2460)$, there are another parameter $\Lambda$ involved by the form factor in the amplitudes, which is also of the order of 1 GeV. Here, we take several typical values for $\Lambda$, which are 1.6, 1.8 and 2.0 GeV, respectively~\cite{Yue:2022mnf}. With the above preparations, we can evaluate the decay widths of $T_{c\bar{s}0}^{+}(2900) \to D_{s1}^+(2460) \pi$ and $T_{c\bar{s}0}^{+}(2900) \to D_{s1}^+(2460) \gamma$ in different frames, which are the $D_{s1}^+(2460)$ molecular frame and the $D_{s1}^+(2460)$ charmed strange meson frame, respectively.

In Fig.~\ref{Fig:Pi}, we present the decay width of $T_{c\bar{s}0}^+ (2900)\to D_{s1}^+(2460)\pi^0$ in the $D_{s1}^+(2460)$ molecular scenario and in the $D_{s1}^+(2460)$ charmed strange meson frame, respectively. From our estimation, one can find the transition width in the $D_{s1}^{+}(2460)$ molecular frame are much larger than that in the $D_{s1}^{+}(2460)$ charmed-strange meson frame. It should be noted that in the $D_{s1}(2460)$ molecular frame, $D_{s1}(2460)$ is composed of $D^\ast $ and $K$, thus, the coupling between $D_{s1}(2460)$ and  $D^\ast K$ is much stronger than that in the charmed strange meson frame. In addition, the kaon is considered as an ordinary meson in the $D_{s1}(2460)$ molecular frame, and the coupling between $D_{s1}(2460)$ and $D^\ast K$ is $S$ wave, which is momentum independent. However, in the $D_{s1}(2460)$ charmed strange meson frame, the kaon is considered as a chiral particle, thus, the $D_{s1}D^\ast K$ vertex is momentum dependent as shown in  Eq. ~(\ref{Eq:LagD1DStarP}) due to the chiral symmetry, which further suppress the width of $T_{c\bar{s}}(2900) \to D_{s1}(2460) \pi$ in the  charmed strange meson frame.

In Ref.~\cite{Yue:2022mnf}, we have investigated the decay properties of $T_{c\bar{s}0}(2900)$ in the $D^\ast K^\ast$ molecular frame. The decay widths of $DK,\ D_s \pi, \ D_s^\ast \rho,\ D_{s1}(2460) \pi, \ D_{s1}(2536) \pi $ and $D^\ast K \pi$ channels were estimated~\cite{Yue:2022mnf}, where the $D_{s1}(2460)$ is considered as a charmed strange meson. For simplify, we just quote the results of Ref.~\cite{Yue:2022mnf} in the following discussions. By comparing the estimated total width with the one reported by the LHCb Collaboration, we obtained the proper $\Lambda_M$ ranges for different $\Lambda$, which is collected in Table~\ref{Tab:Width}. In these parameter range, the estimated widths for $D_{s1}(2460) \pi$ and $DK$ channels are also listed. The ratio of $\Gamma_{D_{s1} \pi}$ and $\Gamma_{DK}$ is estimated to be,
\begin{eqnarray}
\frac{\Gamma_{D_{s1} \pi}}{\Gamma_{DK}} = 0.10\sim 0.14, \label{Eq:Ratio1}
\end{eqnarray}
which is very weakly dependent on the model parameters $\Lambda$ and $\Lambda_M$.

Moreover, one can investigate the decay properties of $T_{c\bar{s}0}^{+}(2900)$ in the $D_{s1}^{+}(2460)$ molecular frame. Considering the isospin symmetry, the strong decay behaviors of $T_{c\bar{s}0}^+(2900)$ and $T_{c\bar{s}0}^0(2900)$ should be the same. When we investigate the decay properties of the  $T_{c\bar{s}0}^+(2900)$ in the $D_{s1}^{+}(2460)$ molecular frame, the partial widths of other decay channels without $D_{s1}^{+}(2460)$ should be the same as those in Ref.~\cite{Yue:2022gym}. Together with the width of $T_{c\bar{s}0}^{+}(2900)\to D_{s1}^{+}(2460) \pi^0$ estimated in the present work, we can roughly obtain the total  width of  $T_{c\bar{s}0}^{+}(2900)$.  Along the same line~\cite{Yue:2022mnf}, one can determine the value of $\Lambda_M$ by reproducing the width of $T_{c\bar{s}0}^{+}(2900)\to D_{s1}^{+} \pi^{0}$, which are also listed in Table~\ref{Tab:Width}. From the table, one can find that the determined range of $\Lambda_M$ in the $D_{s1}^{+}(2460)$ molecular frame is smaller than the one in the $D_{s1}^{+}(2460)$ charmed strange meson frame~\cite{Yue:2022mnf}. In these model parameter, one can find that the partial widths of $D_{s1}(2460) \pi$ and $DK$ channels are in the same order, and the ratio of $\Gamma_{Ds1\pi}$ and $\Gamma_{DK}$ is estimated to be,
\begin{eqnarray}
\frac{\Gamma_{D_{s1} \pi}}{\Gamma_{DK}} = 0.67\sim 1.33.\label{Eq:Ratio2}
\end{eqnarray}
By analyzing Eqs.~(\ref{Eq:Ratio1}) and (\ref{Eq:Ratio2}), one can find the ratio of the widths of  $T_{c\bar{s}0}^{+}(2900)\to D_{s1}^{+} \pi^{0}$ and $T_{c\bar{s}0}^{+}(2900)\to DK$ significantly different across different $D_{s1}^{+}(2460)$ frame work. Thus, this ratio can serve as a valuable tool for evaluating the internal structure of the $D_{s1}^{+}(2460)$.

In addition to $D_{s1} \pi$ channel, we also estimate the widths of the radiative transition from $T_{c\bar{s}0}^{+}(2900)$ to $D_{s1}^{+}(2460)$ in different scenario, and the estimated transition widths depending on the model parameters are presented in Fig.~\ref{Fig:Gamma}. Similar to the case of $D_{s1}(2460) \pi$, our estimations indicate that the widths of $T_{c\bar{s}0}^{+}(2900) \to D_{s1}^{+}(2460) \gamma$ are about 1 keV and more than 50 keV in the $D_{s1}^{+}(2460)$ charmed-strange meson frame and $D_{s1}^{+}(2460)$ molecular frame, respectively.

\section{Summary}
\label{Sec:Summary}
Stimulated by the observation of $T_{c\bar{s}0}(2900)$ by the LHCb Collaboration\cite{LHCb:2022xob, LHCb:2022bkt}, theorist have proposed some different exotic interpretations, among which the $D^\ast K^\ast$ molecular interpretation is the most promising one. Together with $D_{s0}^{*+}(2317)$, and $D_{s1}^{+}(2460)$, the newly observed $T_{c\bar{s}0}(2900)$ makes the states around the $D^{(\ast)} K^{(\ast)}$ thresholds abundant. Different with $T_{c\bar{s}0}(2900)$, the $D_{s0}^{+}(2317)$ and $D_{s1}^{+}(2460)$ could also be the candidates of $P$-wave charmed-strange mesons.  

In the present work, we investigate the pionic and radiative transitions from the $T_{c\bar{s}0}^{+}(2900)$ to $D_{s1}^{+}(2460)$ in the $D_{s1}^{+}(2460)$ charmed-strange meson frame and the $D_{s1}^{+}(2460)$ molecular scenario, respectively.  Our estimations indicate the ratio of the widths of  $T_{c\bar{s}0}^{+}(2900)\to D_{s1}^{+} \pi^{0}$ and $T_{c\bar{s}0}^{+}(2900)\to DK$ are rather different in two different frame. Particularly, the ratio is estimated to be around 0.1 in the $D_{s1}^{+}(2460)$ charmed-strange frame, while the lower limit of this ratio is 0.67 in the molecular frame. Thus, we suggest that the ratio could be employed as a tool for testing the nature of the $D_{s1}^{+}(2460)$. Moreover, our estimation also find the radiative transition width estimated in the $D_{s1}^{+}(2460)$ molecular frame is much larger than the one estimated in the $D_{s1}^{+}(2460)$ charmed-strange meson frame.

%%%

\bigskip
\noindent
\begin{center}
	{\bf ACKNOWLEDGEMENTS}\\
\end{center}
This work is supported by the National Natural Science Foundation of China under the Grant No.  12175037 and 11775050.

\end{document}